# Chromatic aberrations correction of attosecond high-order harmonic beams by flat-top spatial shaping of the fundamental beam

K. Veyrinas[1*], M. Plach[2*], J. Peschel[2], M. Hoflund[2], F. Catoire[1], C. Valentin[1], P. Smorenburg[3], H. Dacasa[2], S. Maclot[2#], C. Guo[2], H. Wikmark[2], A. Zaïr[4], V. Strelkov[5,6], C. Picot[7], C. Arnold[2], P. Eng-Johnsson[2], A. L'Huillier[2], E. Mével[1] and E. Constant[7]

[1]Centre Lasers Intenses et Applications (CELIA), Université de Bordeaux-CNRS-CEA, 33405 Talence Cedex, France
[2]Department of Physics, Lund University, SE-221 00 Lund, Sweden
[3]ASML Research, ASML Netherlands B.V., 5504 DR Veldhoven, The Netherlands
[4]King's College London, Department of Physics, Attosecond Physics Laboratory, Strand WC2R 2LS London, United Kingdom
[5]Prokhorov General Physics Institute of the Russian Academy of Sciences, 38, Vavilova Street, Moscow 119991, Russia.
[6]Moscow Institute of Physics and Technology (State University), 141700 Dolgoprudny, Moscow Region, Russia
[7]Université Claude Bernard Lyon 1, CNRS, Institut Lumière Matière (iLM), F-69622, Villeurbanne, France

* These authors contributed equally to this work.
# Present address: Institut Lumière Matière (iLM), Université Claude Bernard Lyon 1, CNRS, F-69622, Villeurbanne, France

E-mail: eric.constant@univ-lyon1.fr, marius.plach@fysik.lth.se

**Abstract**

Attosecond pulses created by high-order harmonic generation in gases often exhibit strong chromatic aberrations, arising from the broad bandwidth and wavelength-dependent nonlinear light-matter interaction. When the driving laser intensity varies spatially, as for Gaussian driving beams, the apparent source position of the harmonics differs significantly from one order to the next, thus affecting the achievable intensity and duration of the attosecond pulses when they are focused on a target. We show that these chromatic aberrations can be reduced by spatially shaping the fundamental beam to generate high-order harmonics with a driver having a flat-top profile inside the gas medium. By measuring both the intensity profile and wavefront for each harmonic in a plane, we access the extreme ultra-violet (XUV) beam properties and investigate these properties near focus. We observe that controlling chromatic aberrations by flat-top spatial shaping strongly reduces the variation of the XUV spectrum on the beam axis during propagation and, in return, the longitudinal sensitivity of both the temporal profiles and the temporal shifts of the focused attosecond pulses.

## 1. Introduction

High-order harmonic generation (HHG) in gases is a source of phase-locked broadband extreme ultra-violet (XUV) pulses that are now commonly used in applications requiring femtosecond and/or attosecond resolution [1], [2], [3], [4]. Attosecond dynamics is for instance accessible via XUV – Infrared (IR) [5], [6] or XUV-XUV pump-probe experiments [7], [8]. XUV-IR experiments use either a single attosecond pulse (streaking technique) [10] or a train of pulses (Reconstruction of Attosecond Beating by two-photon Transition, or RABBIT technique)[11]. Since both approaches are mainly based on phase variation measurements of an oscillatory signal, the spatial properties of the XUV beams are not crucial [12]. Analyses are indeed often performed by averaging over space, thus neglecting the influence of any possible spatio-dependent effect such as chromatic aberrations. Attosecond XUV-XUV pump-probe experiments, on the other hand, require high focused intensities. Hence, focusing XUV beams to small spots while maintaining their attosecond temporal structure is crucial [13], [14], [15], [16], [17]. These properties are also essential in high resolution imaging using broadband XUV radiation [18].

To achieve intense attosecond pulses on target, it is necessary to focus all frequency components at the same position. This requires all harmonics to have similar spatial properties, which is often not the case due to intrinsic chromatic aberrations [19], [20]. The origin of the chromatic aberrations lies in the fact that harmonic dipole phases [21], inherent to HHG, is dependent upon the interplay between the harmonic order and laser intensity [22]. Furthermore, the dispersive generating medium can have an index that varies with space via the laser intensity and medium ionization yield. In the generating medium, the phase of the emitted XUV radiations is therefore space and wavelength-dependent and evolves radially when the laser intensity presents a radial dependence. With Gaussian beams, the radial laser intensity variations in the generating medium induces a wavefront curvature that changes with harmonic order and causes chromatic aberrations [19] , [20], [23].

Order-dependent far field spatial profiles have been observed in many different generating media, e. g., jets [24], cells [25], [26], semi-infinite cells and filaments [27], [28] or gas filled capillary [29], [30]. In addition, experimental measurements have shown evidence that harmonics originate from different source points which depend strongly on the process order [19], [20], [31], [32], [23], [26]. When refocused, the harmonics are therefore focused on different, longitudinally separated, positions. As a consequence, the attosecond temporal profile varies along the propagation axis [33], [34], [24], [26]. A scientific effort is therefore increasingly devoted to exploring and controlling the spatial properties of high-order harmonics and attosecond pulses [34], [35], [24], [36], [37], [19], [20], [31], [32], [38], [39], [26], [23], [40] which is the aim of the study presented here.

In this work, we generate high-order harmonics with an IR beam having a flat-top profile near focus [41], [42], [43]. While flat-top beams can be obtained from Gaussian beams after propagation in long media [44] when ionization induces intensity shaping, we chose, here, to shape the flat-top beam directly with a phase mask to disentangle shaping and propagation effects. The beam-shaping apparatus is robust, stable, and versatile and allows us to achieve beam profiles that can be super-Gaussian, flat-top or annular, fine-tuned around the flat-top configuration by opening or closing an iris. With flat-top shaping and a short medium, the transverse intensity gradient is reduced in most of the generating volume and the harmonic properties become less dependent on the generating conditions than with a Gaussian fundamental beam. After refocusing the harmonics, we characterize the XUV wavefront curvature and spatial profile with the Spectral Wavefront Optical Reconstruction by Diffraction (SWORD) technique [45] and thereby measure the position and sizes of the XUV focus for each harmonic generated in a gas jet (see supplementary material, SM, for a gas cell). We observe that the harmonic beams generated with a flat-top shaped fundamental beam are spatially much narrower in the far field [42] than with the Gaussian beam which implies that the XUV foci (or apparent sources) are larger. This has a large impact on the attosecond XUV beam quality as we find that the different harmonics can be focused much closer to each other (relative to the XUV confocal parameter) using a flat-top driving field as compared to the standard generation with Gaussian beams. Chromatic aberrations can therefore be controlled thus improving the spatio-temporal characteristics of the attosecond pulses. These results are compared to numerical simulations consistent with our observations.

## 2. Experimental method

The experimental setup is schematically shown in Fig. 1. High-order harmonics are generated at 10 Hz repetition rate in a gas medium (jet or cell) with a 40 fs high-energy (up to 45 mJ after compression) titanium sapphire IR laser driver centered at $\lambda$ = 808 nm that is spatially filtered and wavefront-corrected. Wavefront control is performed with a deformable mirror located under vacuum. The IR beam with waist W = 27 mm, is truncated by a motorized iris and focused with an f = 8.7 m focal length mirror in a gas medium (pulsed gas jet with 250 µm jet nozzle and 5 bar backing pressure or 1 cm long gas cell). The IR focus position vs the gas medium is adjusted by controlling the curvature of the deformable mirror [26]. The IR beam shape at focus is observed with a camera located at a position that mimics the gas medium position. This observation is performed in air with an attenuated beam that is transmitted through a folding mirror (Fig. 1).

A phase mask can be inserted in the path of the fundamental beam to achieve spatial shaping near the IR focus. The mask is an anti-reflection coated, 3 mm thick, $SiO_2$ plate (see Fig. 1) with an additional 880nm thick, 20 mm diameter central area. The thickness of the $SiO_2$ central part is chosen to create a $\pi$ dephasing between the central and outer parts of the IR beam. It induces destructive interferences at focus on the beam axis between the inner and outer beams when both are focused [42], [46]. These interferences redistribute light away from the axis and leads to a flat-top profile in the radial dimension under proper conditions. When the iris is closed to a diameter of 20 mm, the phase mask has no shaping effect, and the IR beam is a truncated Gaussian beam. The beam size at focus is then approximately 300 µm FWHM. Furthermore, this iris diameter is sufficiently large to provide optimum conditions for HHG with a regular apertured Gaussian beam. When the iris is opened to larger diameters, shaping occurs leading to a near flat-top beam at IR focus as detailed in the following.

The generated harmonics are reflected by an $SiO_2$ plate that attenuates the IR, filtered by an Al foil and reflected by two toroidal mirrors placed in a Wolter configuration focusing the XUV beam and providing a 35-fold demagnification of the harmonic source [47]. The XUV focus is located at the entrance of a slit-less flat-field spectrometer that consists in a variable line spacing Hitachi grating and an MCP detector imaged with a CCD camera and enables for the characterization of the full beam.

An additional 120 µm wide slit, perpendicular to the grating grooves and mobile in the direction of the grooves, can be inserted between the XUV focus and the spectrometer grating to select a small portion of the beam. The transmitted beam hits the MCP at a position that depends on the XUV radiation wave vector at the slit position. Observing the impact position as a function of the slit position provides the radial evolution of the wave vector orientation and thereby the radius of curvature of the XUV beam in the slit plane. The XUV beam intensity profile in this plane is also measured by integrating the transmitted XUV signal as a function of the slit position. This SWORD measurement [45], performed for each harmonic, provides the radial intensity profile and wavefront curvature in the slit plane for each harmonic.

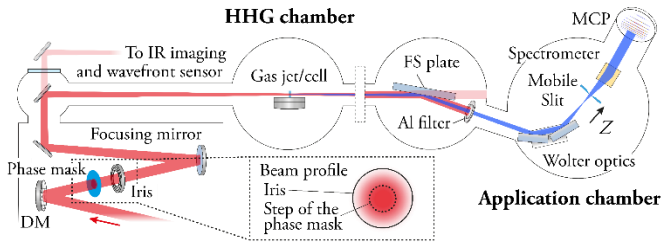

*Fig. 1. Experimental setup used for high-order harmonic generation and characterization. The spatially filtered IR laser beam is wavefront corrected by a deformable mirror (DM) and truncated by a motorized iris before focusing on a gas target. A phase mask (anti-reflection coated $SiO_2$ plate with an additional 20 mm diameter, 880 nm thick $SiO_2$ step on the central part) can be inserted to spatially shape the IR beam near its focus. Harmonics are generated in a gas jet located at the IR focus (for gas cell see supplementary material, SM). The XUV beam is then filtered spectrally by a fused silica (FS) plate and a 200 nm thick Al filter before being refocused by Wolter optics at the entrance of a slit-less flat-field spectrometer. An additional slit can be translated in the direction of the spectrometer grating grooves to perform a SWORD measurement, thus providing the wavefront and intensity profiles of each harmonic beam in the slit plane. Z is the longitudinal propagation coordinate in the application chamber and Z = 0 is arbitrarily set at the focus position of H15.*

Shaping of the fundamental beam is achieved with the phase plate and controlled by a motorized iris. Fig. 2 shows a cut of the shaped IR intensity profiles at focus as a function of the iris diameter. When the iris is closed to 20 mm diameter (lower curve) the IR intensity profile at focus resembles a Gaussian profile. This configuration represents our reference Gaussian configuration in the following. When the iris is opened to 27 mm diameter, the beam exhibits an annular shape with a local intensity minimum in the center of the beam. For intermediate iris diameters, typically between 24 and 26 mm, the beam is shaped to a flat-top or super-Gaussian profile. This shaping arises near focus from on-axis destructive interferences between the inner and outer parts of the beam that are very dependent on the exact beam shape and on the centering of the plate. It is therefore difficult to obtain a perfect symmetry, but the observed beam profile nicely follows our previous simulations [46] (see SM fig. S2). This approach provides a reference Gaussian shape for $\phi_{iris}$ = 20 mm and near flat-top beams for $\phi_{iris} \geq 24$ mm.

The beam size (FWHM) at focus, increases with the iris diameter and changes by a factor of approximately 2 between the reference case ($\phi_{iris}$ = 20 mm, FWHM = 300 µm) and the flat-top beam (FWHM = 500 to 540 µm). The pulse energy increases slightly when the iris is opened. We measure a reduction of the focused laser intensity due to beam shaping by a factor of ~ 3.

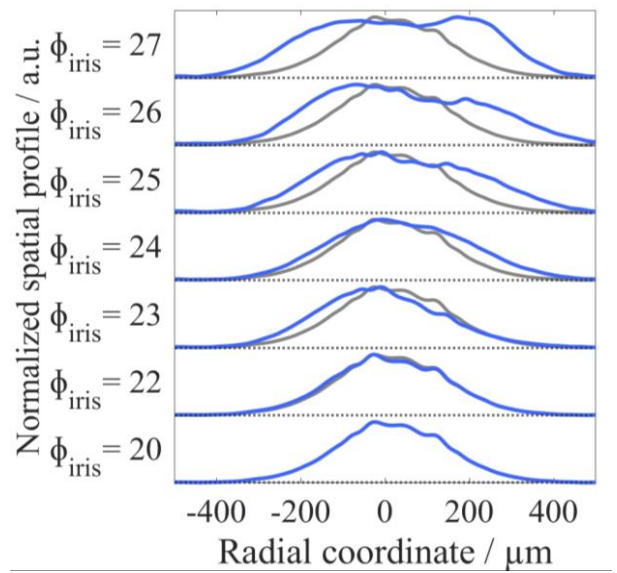

*Fig. 2. Normalized cuts of the IR beam intensity profile at focus as a function of the iris diameter ranging from 20 mm to 27 mm. The reference Gaussian beam (thin grey line) is obtained when the iris is closed to 20 mm diameter in which case the beam is not shaped by the phase mask*

## 3. Results

### 3.1 Experimental results

Harmonics are generated in argon with both flat-top and Gaussian beams. The IR intensity decreases with the shaping and only harmonics 11 (noted H11) to 19 (H19) are observed with the flat-top beam while harmonics with higher orders are easily obtained with the truncated Gaussian beam. In the following, the same laser energies are used with flat-top and Gaussian beams and only harmonics that are generated in both configurations (H11 to H19) are considered.

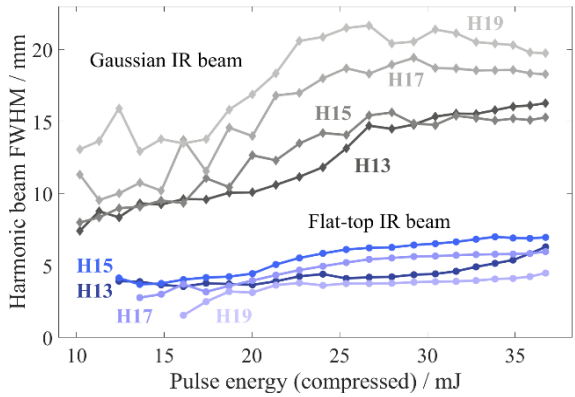

*Fig. 3. XUV beam size (FWHM) on the MCP detector for harmonic orders 13 to 19 for several laser energies with the flat-top spatial shaping ($\phi_{iris}$ = 25 mm, circles, blue curves) and the reference Gaussian beam ($\phi_{iris}$ = 20 mm, diamond markers, grey curves). The gas jet is located at the IR focus.*

Fig. 3 shows that harmonics generated with a Gaussian beam (grey curves, diamond markers) have a size increasing with the harmonic order at a given intensity. This is typical when harmonics generated at focus via the short quantum path are detected. It can also be observed that the XUV beam sizes increase with the intensity. When harmonics are generated with a flat – top shaped beam (blue curves), the XUV beams are smaller by approximately a factor of two [42], [43], [48] and, opposite to the Gaussian beam case, the beam size decreases with the harmonic order at a given intensity. The evolution of the XUV beam size with intensity is also less pronounced with the flat-top than with the Gaussian beam. Since the flat-top spatial shaping is achieved near the IR focus on a limited longitudinal range [46], these observations are performed with the jet located at the IR focus. It is however known that with Gaussian beams, the XUV beam divergence changes with the generating medium position. In general, locating the medium at the focus of the Gaussian fundamental beam does not lead to minimum divergence [19], [20], [49]. We therefore measured the divergence of the XUV beam generated with the Gaussian beam for several longitudinal positions of the jet relative to the IR focus. The XUV divergence remained larger than that observed with the flat-top driver and we observed that the positions of minimum divergence changes with harmonic order [19], [20].

In the shaped beam case, we observed similar beam sizes for all harmonics. This indicates that the impact of the order dependent spatial phase variation is reduced by the shaping. This is expected as the atomic phase, $\phi_q(r)$, depends on the intensity and on the harmonic order but its variations are reduced when the intensity is independent on r, the radial coordinate. Flat-top spatial shaping also reduces the radial evolution of the phase term, $\phi_0(r)$, accumulated by the fundamental during propagation in a partially ionized medium as ionization is also independent on the radial distance with a flat-top beam. This reduces plasma-induced defocusing that is also known to affect the divergence of harmonics [50]. At the low intensity used here for the flat-top beam, ionization must be very limited. For the Gaussian case, the IR intensity is higher, and simulations show that ionization affects the XUV beams (see supplementary material, SM).

These observations are corroborated by a spectrally integrated direct observation of the XUV beam on an X-ray CCD camera (Andor). We systematically observe that the XUV beams are smaller when generated with a flat-top beam than when generated with the reference Gaussian beam (see SM). These observations show that the flat-top shaping is a way to minimize the influence of the intensity dependent spatial phase variation on the harmonic beam divergence.

The SWORD measurements show also noticeable differences between the wavefront radii of curvature of the harmonics generated by flat-top and Gaussian beams. Fig. 4 presents the outcome of the measurements performed with the Gaussian and flat-top IR beams when harmonics are generated in a gas jet (see SM for gas cell, Fig. S5). The radii of curvature, measured in the slit plane, are on the order of the geometrical distance between the Wolter focus and the slit (approximately 12 cm) but their evolution with harmonic order is different. Harmonics generated with a Gaussian beam exhibit radii of curvature, $R_q$, that decrease with increasing harmonic order, q. In contrast, when harmonics are generated with a flat-top beam, $R_q$ increases with q. Measurements performed with a gas cell show similar trends (see SM Fig. S5.).

The exact values of $R_q$ extracted from these measurements strongly depend on the calibration and the accuracy of the measurement, limited by the spatial resolution on the MCP.

However, the relative evolution of $R_q$ with harmonic order is less sensitive to calibration and we estimate that a relative error of less than 1% is achieved. The error bars can also be directly estimated from a comparison between the first and second orders of diffraction of the XUV grating which should give the same radii of curvature. Indeed, observed differences are less than 0.5% of the radii of curvature.

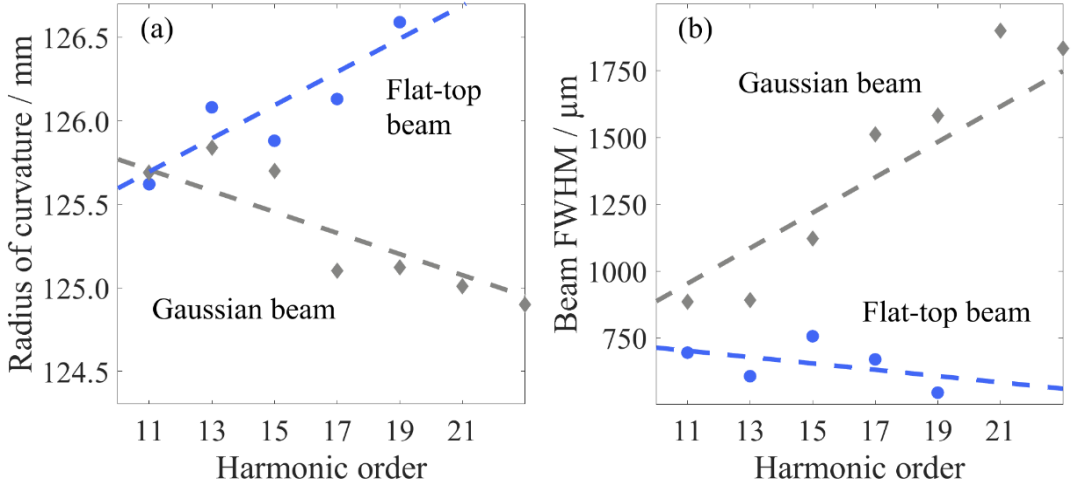

*Fig. 4. (a) Radii of curvature, $R_q$, and (b) beam sizes (FWHM) of the harmonics generated in a gas jet with a Gaussian beam (grey diamond markers) or with a shaped flat-top beam (blue circle symbols). Harmonics are generated in a gas jet. The measurements are performed in the slit plane located approximately 12 cm after the Wolter focus. The results obtained with a 1 cm gas cell show a similar trend (see SM). Dashed lines are a linear fit representing a guide for the eye.*

### 3.2 Analysis within Gaussian approach

We use these measurements to estimate the positions of the harmonic foci. Harmonic beams have profiles that are close to Gaussian (see SM Fig. S3, S10 and S11) and we use the Gaussian approach presented in Quintard *et al.* [19] to extract the positions and sizes of the harmonic foci (a similar approach is presented in Wikmark *et al.* [20]). From the positions and sizes of the foci, the spatial properties of the harmonic beam (beam size and radius of curvature) can be determined in any plane. Our model relies on the assumption that the XUV beams are ideal Gaussian beams with quadratic wavefront which correspond well to our observations. This approach also assumes that the $M^2$ factor of each harmonic beam is equal to 1 which is not measured here but provides a good fit of the experimental data (see SM figures S10 and S11). Under these assumptions, the size $W_{0q}$ and position $Z_{0q}$ of the harmonic waist with respect to the slit position are uniquely defined from the measurement of $R_q$, the radius of curvature (Fig. 4 (a)) and $W_q$, the beam size in the slit plane (Fig. 4 (b)). With $\lambda_q$ the wavelength of the harmonic, we have:

$$W_{0q} = \sqrt{\frac{\left[\frac{R_q \lambda_q}{\pi W_q}\right]^2}{1+\left[\frac{R_q \lambda_q}{\pi W_q^2}\right]^2}} \quad \text{Eq. (1)}$$

$$Z_{0q} = -\frac{R_q}{1+\left[\frac{R_q \lambda_q}{\pi W_q^2}\right]^2} \quad \text{Eq. (2)}$$

The foci sizes and positions, represented in Fig. 5 (a) and (b), change with harmonic order. Positions are represented with reference to the 15[th] harmonic focus. The XUV foci sizes obtained with a Gaussian fundamental beam are found to be in the range of 2 to 3 µm, in agreement with former measurements on this source [26], [47], [51]. In the case of a flat-top fundamental beam, the waist sizes are larger and reach 4 to 5 µm. From these values, we can estimate the beam size in any plane. Those estimated in the plane of the MCP agree well with the measured beam profiles (see SM, Fig. S10 and S11) which further validates our approach.

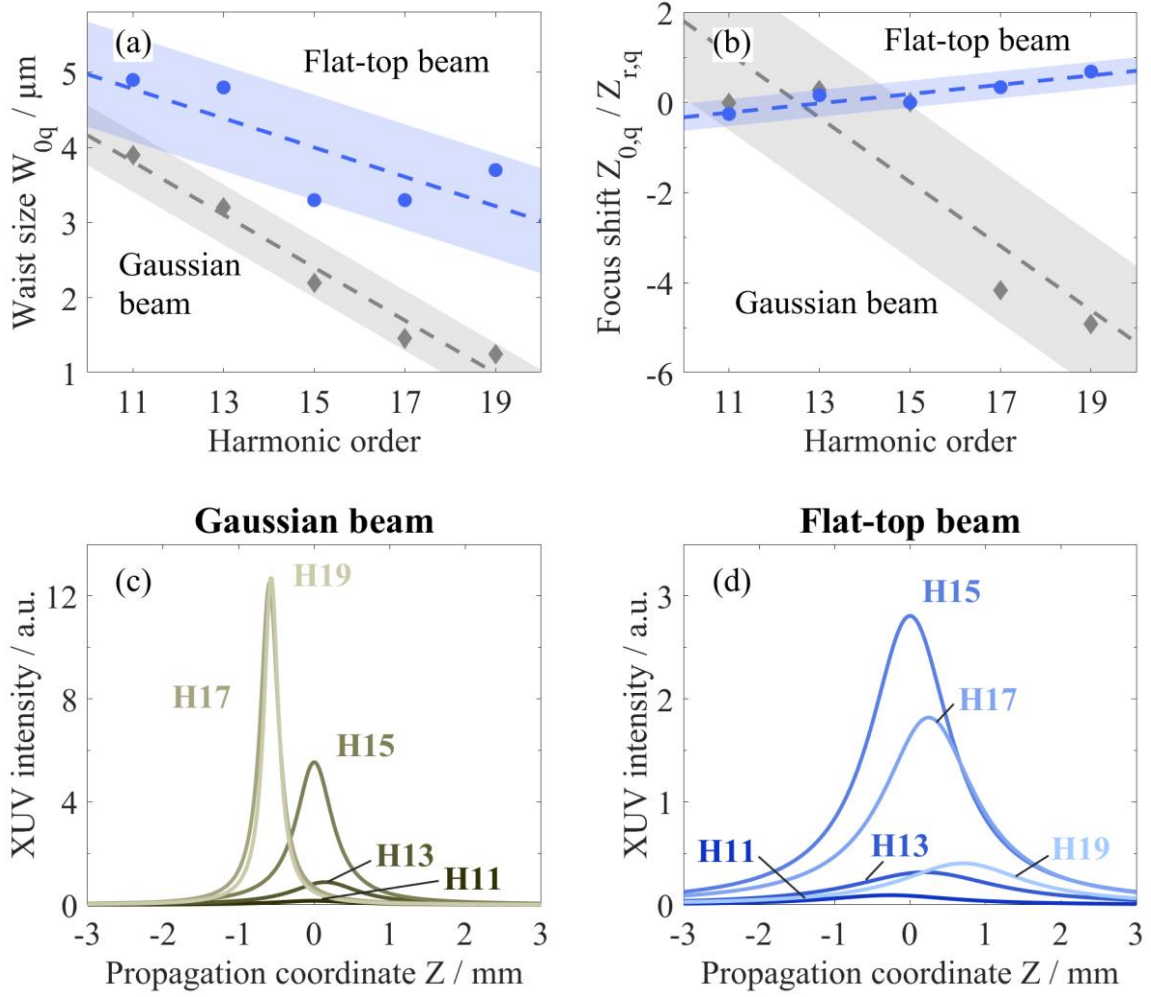

*Fig. 5. (a) Waist size and (b) focus shift, $Z_{0,q}$, divided by the Rayleigh range of each harmonic, $Z_{r,q}$, as a function of harmonic order (blue: flat-top, grey: Gaussian) simulated with the measured beam sizes and radii of curvatures. Subfigures c) and d) illustrate how the harmonics are focused. It shows the simulated on-axis XUV intensity for the harmonic orders 11 to 19 for the fundamental Gaussian beam (c) and for the flat-top beam (d) respectively. Z = 0 is arbitrarily set to the position of the focus of harmonic 15.*

Fig. 5 (c) and (d) show the harmonic intensity along the propagation axis when harmonics are generated in a gas jet with the reference Gaussian beam (c) and with a flat-top beam (d) with the experimentally measured characteristics. The harmonic foci are separated longitudinally in both cases, but the separation is smaller than the XUV confocal parameters for the flat-top driver while it is larger for the Gaussian beam. Fig. 6 shows the average XUV photon energy along the propagation axis and the XUV bandwidth. The bandwidth is estimated to be 2.35 σ (FWHM = 2.35σ for a Gaussian distribution) where σ is the root mean square difference of the photon energy distribution. Far from focus, the on-axis bandwidth is constant and does not evolve significantly with propagation. Near the focus, we observe a change of the mean photon energy and of the XUV bandwidth. With the Gaussian beam, the bandwidth changes by almost a factor 2 (from 4.8 to 8 eV) with longitudinal position while for the flat-top beam it changes only by 26 % (5.4 to 7.3 eV). In both cases, the on-axis bandwidth decreases near focus.

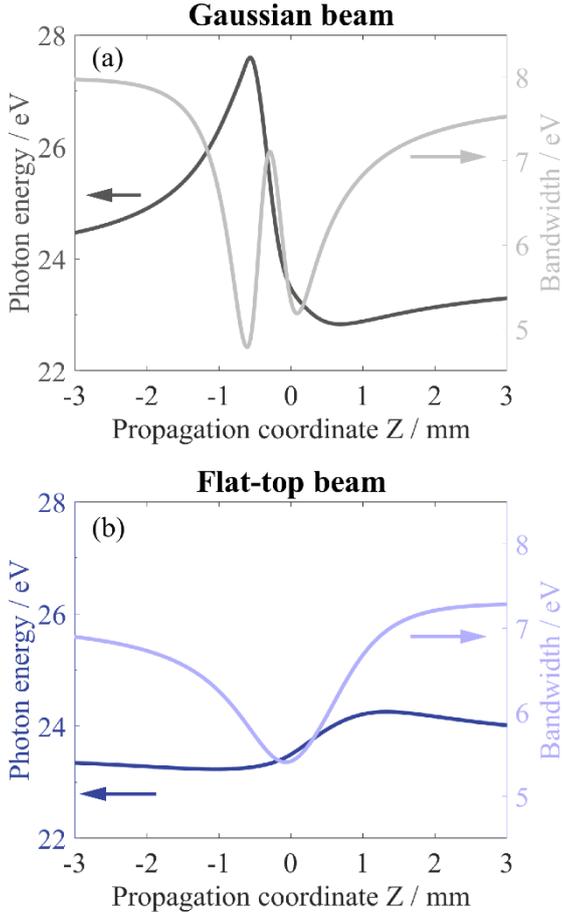

*Fig. 6. Evolution of the average photon energy (thick line) and bandwidth (light line) for (a) Gaussian and (b) flat-top shaped fundamental beams*

These observations show that chromatic aberrations are present in XUV harmonic beams and are reduced when the generating beam has a flat-top shape in the generating medium. Chromatic aberrations impact the focusing and local bandwidth and consequently the attosecond pulse structure as illustrated in the following.

The duration of attosecond pulses depends on the bandwidth and on the dephasing between its frequency components ("atto chirp") [52]. Measuring the spatial characteristics of each harmonic beam and their relative amplitude allows us to estimate the temporal profile along the propagation axis using the measured harmonic amplitudes and including the Gouy dephasing. We assume that all harmonics are in phase at infinity and present no attochirp (Fourier limited pulses). This represents the ideal case where the attosecond pulse duration is the shortest compatible with its spectrum. We observe that the pulse duration changes near the focus by a large fraction of its asymptotic value when the harmonics are generated with a Gaussian beam (here 31% with estimated durations between $\tau_{asympt} = 280$ as and $\tau_{max} =$ 362 as) while the typical variation is of the order of 10 to 20 % for a flat-top shaped fundamental beam (here 16 % with estimated durations between $\tau_{asympt} = 290$ as and $\tau_{max} = 336$ as). Similar effects are observed with harmonic generation in a gas cell (see SM, Fig. S6).

The change of pulse temporal profile with longitudinal propagation is illustrated in Fig. 7 which shows the on-axis temporal intensity profile as a function of the propagation coordinate after normalization. This normalization suppresses the on-axis intensity evolution that is due to beam divergence. We observe a significant distortion of the pulse near focus and a pulse duration changing with propagation when harmonics are generated with the Gaussian beam (Fig. 7 (a)). The flat-top shaping of the fundamental beam has a strong spatial smoothing effect on the XUV beam and the pulse duration on axis changes only very slightly with propagation (Fig. 7 (b)).

These results also reveal that the chromatic aberrations impact the relative dephasing between harmonics which evolves with propagation. The Gouy phase shift evolution affects the relative harmonic dephasing and is the strongest near the XUV foci. It affects the timing of the XUV pulse maximum as compared to t = 0 that represents the center of the XUV pulse at asymptotic positions. Near XUV focus, we observe a shift of the center of the pulse by more than 100 as when Gaussian beams are used. This shift reduces to 30 as when a flat-top beam is used. For pump – probe experiment involving XUV pulses and IR fundamental, the observed shifts near focus can affect the temporal resolution.

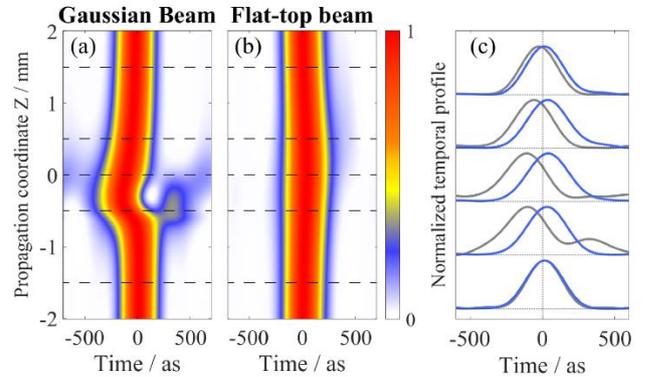

*Fig. 7. Spatial evolution of the attosecond temporal profile for harmonics 11 to 19 emitted in a jet with a Gaussian fundamental beam (a) or a flat-top shaped fundamental beam (b). The corresponding temporal profiles are shown in sub-figure (c) for five longitudinal positions (Z = -1.5 mm, -0.5 mm, 0 mm, 0.5 mm and + 1.5 mm) as indicated by the dashed line through the temporal profile. The blue line is the profile obtained with the flat-top fundamental and the grey line is obtained for the Gaussian beam. The results obtained with a 1 cm gas cell can be found in the SM, Fig. S7.*

3.3 TDSE based simulations

More advanced simulations are performed with accurate atomic response calculations and going beyond the Gaussian model discussed previously. The simulations of the experimental geometry assume the generating IR field being either a truncated Gaussian beam or a truncated Gaussian beam with the phase mask. In the latter case the field at the target has a radial flat-top spatial profile similar to the one shown in Fig 2 (a) with $\phi_{iris}$ = 25 mm. The microscopic response is calculated via 3D Time dependent Schrödinger equation (TDSE) for a model argon atom. The generating medium is approximated by an infinitely thin plane layer. The amplitude and phase of the atomic response are calculated at a set of transverse positions. The IR pulse duration is 40 fs, and the peak IR intensity on axis is $2 \times 10^{14}$ W/cm$^2$ in one set of calculations and $3.5 \times 10^{14}$ W/cm$^2$ in another one (see SM). Finally, the effect of the focusing Wolter optics is also simulated.

The simulated XUV divergence is found to be smaller with the flat-top beam compared to the Gaussian beam, as observed experimentally. Figure 8 shows that order dependent XUV foci shifts are present. They are larger with a Gaussian than with a flat-top driver. For the Gaussian beam, the foci of lower order harmonics are longitudinally shifted downstream with respect to the ones of the high-order harmonics by about 1 mm. This can be observed in the experimental data shown in Fig. 5. The foci longitudinal separation for the Gaussian beam leads to a more pronounced bandwidth and pulse duration evolution along the propagation axis in the range where the harmonics are focused (Fig. 8 (c) and (d)), in agreement with the predictions of the Gaussian model. Contrary to these predictions, however, the distances over which the harmonics are focused are smaller with the flat-top than with the Gaussian. As the XUV divergence is smaller for harmonics generated with the flat-top beam, this is a signature of the poor optical quality for the XUV beam generated with the Gaussian beam. Simulations also show a net asymmetry in the intensity evolution along propagation. The differences between the Gaussian model and the TDSE results illustrate also the range of validity of the (simplified) Gaussian model. For instance, this model cannot reproduce structured XUV foci which are obtained at high intensity in TDSE simulations (see Fig. S13). Also, it does not include plasma-induced defocusing or ionization induced intensity reshaping that can be significant for high intensity and long/dense media [44], [53], [54], [55]. Intensity reshaping due to propagation, which is not monitored in the present work, is expected to be weak as the medium is very thin and the intensity not too high. TDSE simulations show nevertheless that plasma induced defocusing can have an impact at high intensity on the XUV wavefront as it changes the IR wavefront and thereby also the XUV divergence (see SM). In these simulations, the XUV beam profile is found to be more regular when generated with a flat-top driver than with a Gaussian driver.

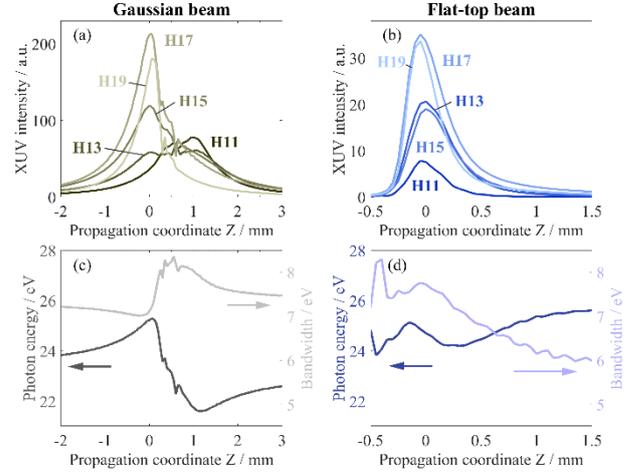

*Fig. 8. Evolution of the XUV intensity on the axis as a function of the longitudinal position for harmonics generated with a truncated Gaussian beam (a) or with a flat-top beam (b). Z = 0 is arbitrarily set to the position of the focus of harmonic 15 (c-d) Corresponding evolution of average photon energy (thick line) and the bandwidth (thin line) on the beam axis as a function of the longitudinal position. These simulations are performed with TDSE (see text).*

## 4. Discussion

The study is performed here on a limited spectral range (H11 to H19) that corresponds to the bandwidth generated by the flat-top beam. Similar chromatic aberrations are observed with Gaussian beams over the full bandwidth (see SM, Fig. S8 and S9) and in general chromatic aberrations increase with the emitted bandwidth. This may become a strong limiting factor to achieving shorter attosecond pulses with very large bandwidth [56], [57]. The work presented in this article shows that shaping the fundamental beam can be an efficient way to reduce chromatic aberrations. More complex spatial shaping can be performed with advanced technology such as spatial light modulator (SLM) [48] opening the possibility to reduce chromatic aberrations over a much larger bandwidth. Furthermore, the study of flat-top HHG at higher intensities is of particular interest to see over which maximum bandwidth the foci shift can be neglected. In return, this would give an estimate of the shortest attosecond pulses that can be used near focus without considering the beam spatial properties. Increasing the intensity for the flat-top IR driver would also increase the XUV flux. This is technically possible and would require the use of shorter focal lengths or higher laser energies.

The chromatic aberrations studied here arise from the fact that inside the generating medium the harmonics have spatial characteristics (wavefront and beam size) evolving with the harmonic order. We observe the same effect in a gas jet and a 1 cm cell (see SM) that are both minor compared to the IR confocal parameter. It would also be interesting to study chromatic aberrations of harmonics generated with longer media or in a guided configuration [58], [59]. In a longer medium, ionization induced reshaping of the IR intensity profile can occur and can also lead to flat-top intensity profile after propagation [44]. It would be particularly interesting to see to which extent both flat-top configurations are equivalent and if both can help controlling the chromatic aberrations. In a guided configuration, the wavefront of the fundamental beam is flat, as it is the case here at the IR focus, but the XUV phases remain dependent on the IR intensity and on the plasma density that both change radially. Even in a guided configuration, the XUV wavefront will therefore not be flat and will depend on the harmonic order if the harmonics are not guided which is usually the case. In fact, when defocusing is neglected and for plateau harmonics that are significantly reabsorbed, there is little difference between HHG in a guided configuration and HHG at focus in a gas medium that is smaller than the IR confocal parameter as performed here. We therefore anticipate that similar chromatic aberrations should also exist in XUV beams generated in a guided configuration and that the observed phenomenon is very general.

**5. Conclusion**

In summary, we show that performing HHG with a flat-top spatially shaped fundamental beam provides control of the XUV beam properties and allows us to reduce XUV chromatic aberrations as compared to the usual case where HHG is performed with a truncated Gaussian beam. The position of each harmonic focus is measured by the SWORD technique showing that the harmonic foci are separated longitudinally. The chromatic aberrations are strong enough to affect the XUV bandwidth locally and the attosecond temporal profiles simulated on axis show a pulse duration that changes significantly during propagation when Gaussian beams are used for HHG. For harmonics generated by a flat-top shaped beam, focus separation still exists but the XUV beam divergence is strongly reduced while the apparent size of the harmonic sources is increased. In this case, the focus offsets are smaller than the XUV Rayleigh length and the impact of XUV chromatic aberrations is reduced. These chromatic aberrations are associated with a strong longitudinal evolution of the XUV bandwidth and of the attosecond temporal profiles when harmonics are generated with a truncated Gaussian beam especially near the XUV foci. The variation of the attosecond pulse duration along the propagation axis and the associated temporal shift are strongly reduced when flat-top spatially shaped fundamental laser beams are used for HHG. This spatial-shaping-induced control will be highly beneficial to study attosecond dynamics with high temporal resolution and to achieve high XUV intensities.

**Acknowledgements**

The research leading to these results has received funding from LASERLAB-EUROPE (grant agreement no. 654148, European Union's Horizon 2020 research and innovation programme). The authors acknowledge support from the Swedish Research Council, the European Research Council (advanced grant QPAP, 884900), the Knut and Alice Wallenberg Foundation and Région Nouvelle-Aquitaine through the 'OFIMAX' project (contract N◦ 184289). AL is partly supported by the Wallenberg Center for Quantum Technology (WACQT) funded by the Knut and Alice Wallenberg foundation. M. P. acknowledges the support of the Helmholtz Foundation through the Helmholtz-Lund International Graduate School (HELIOS, HIRS-0018). V.S. acknowledges support from Theoretical Physics and Mathematics Advancement Foundation "BASIS". We acknowledge the expert assistance from Anders Persson.

**Data availability statement**

The data that support the findings of this study are available upon request from the authors.

Supplementary material:

The spatial profile of the fundamental IR beam was estimated at the focus position by measuring the profile of a small fraction of the beam. Fig. S1 shows the image of the focused truncated Gaussian beam and of the spatially shaped flat-top beam.

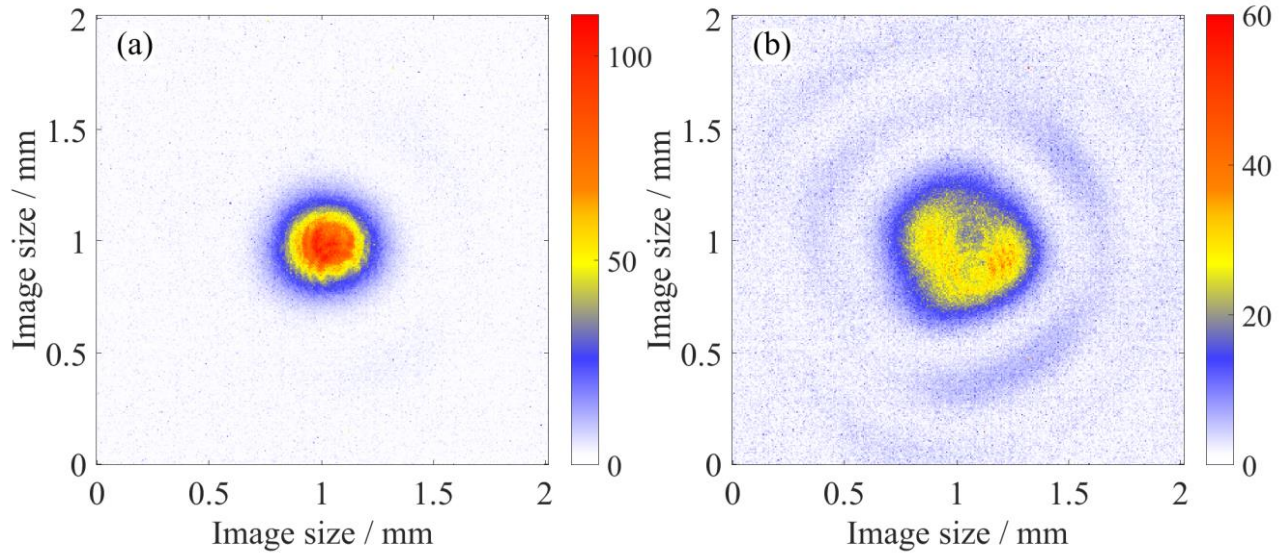

*Fig. S1: Infrared spatial profile at focus (a) without the spatial shaping (incoming Gaussian beam clipped with an iris of diameter $\Phi_{iris}$ = 20 mm) and (b) with the flat-top spatial shaping (phase plate used with an iris of diameter $\Phi_{iris}$ = 27.5 mm).*

The spatial profile and wavefront of the shaped beam was simulated (Fig. S2) from z = - 20 cm to z = + 20 cm (here z = 0 refers to the position of the IR focus obtained without shaping). It shows flat-top intensity profile near z = 0 and an inversion of curvature in this plane showing that the flat-top profile is achieved with this approach with a flat wavefront of the fundamental.

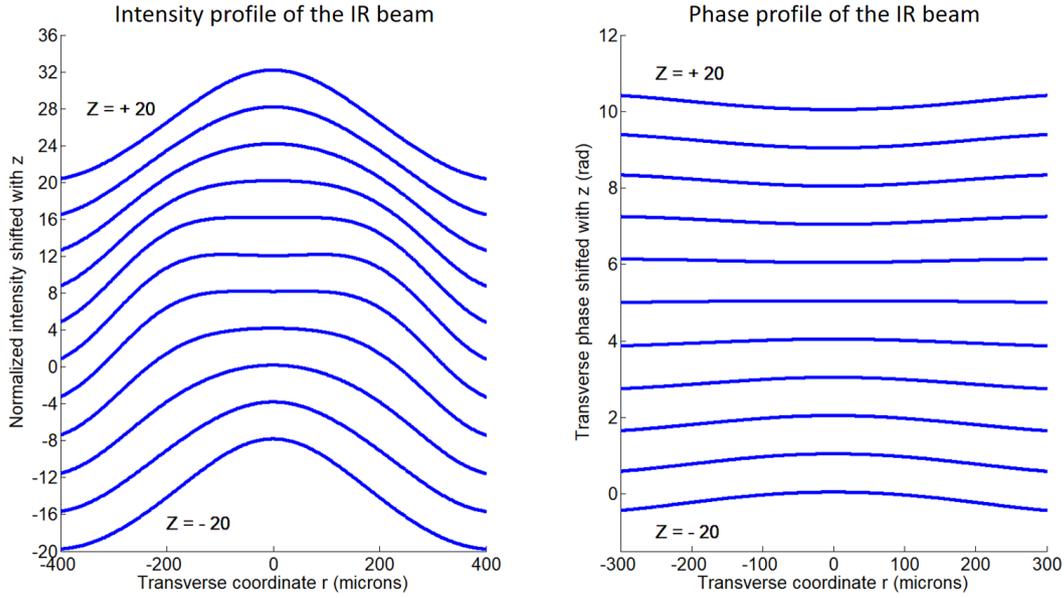

*Fig. S2: Simulated spatial profiles and phase front of the flat-top IR beam shifted as a function of z (z = -20 cm to + 20 cm with 4 cm steps). Simulations performed with W = 27 mm, $\phi_{plate}$ = 20 mm, $\phi_{iris}$ = 25 mm, $\delta\varphi = \pi$. A flat-top beam shape is achieved near z = 0 and the wavefront curvature changes sign around z = 0 (here z = 0 refers to the position of the IR focus obtained without shaping). With these parameters, the IR wavefront is flat near the flat-top position.*

The XUV spatial profiles are also directly measured with an XUV camera (Fig. S3) in which case, all harmonics transmitted by the Al filter are detected simultaneously. These images, obtained in the far field, show that the XUV beams are smaller when generated with the flat-top fundamental beam than when generated with the truncated Gaussian beam.

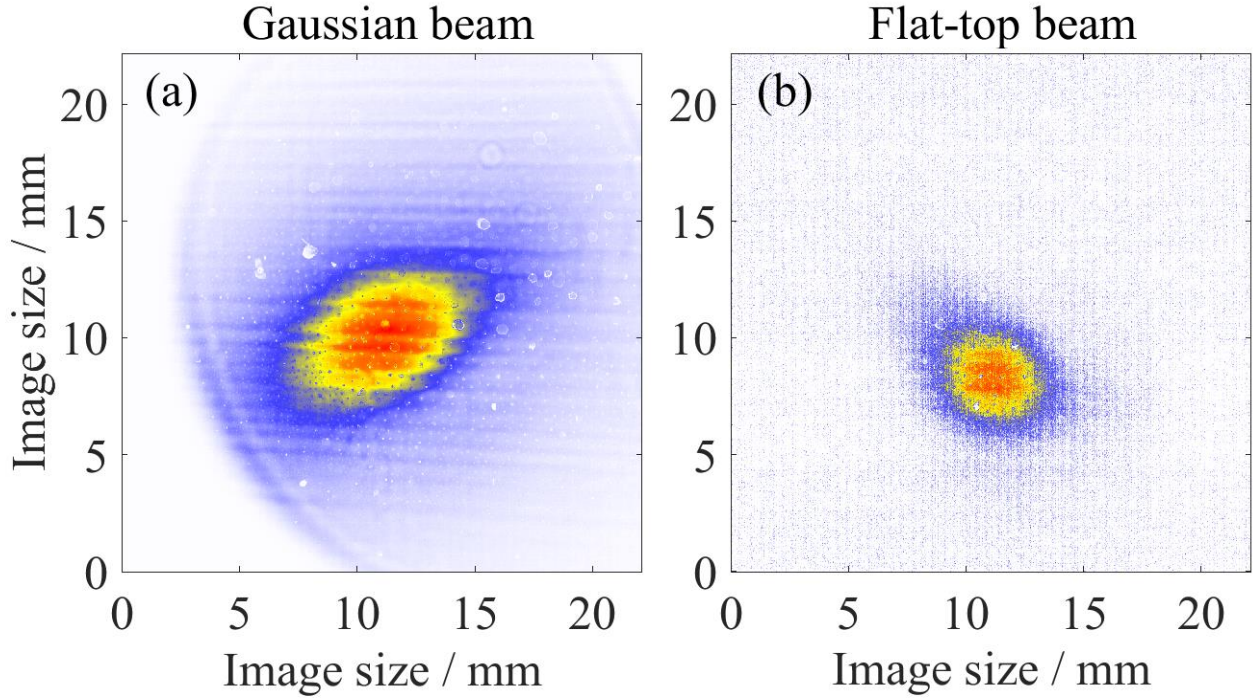

*Fig. S3: Footprint of the XUV beam detected with an Andor back illuminated CCD camera. (a) Harmonics generated with a Gaussian beam, (b) Harmonics generated with a flat-top beam.*

The measurements of the radii of curvature are also performed after varying parameters that are crucial in HHG such as iris diameter and pulse energy. The trends discussed in the paper are robust and reproduced under many different conditions both with the gas jet (Fig. S4(a)) and with the gas cell (Fig. S4(b)).

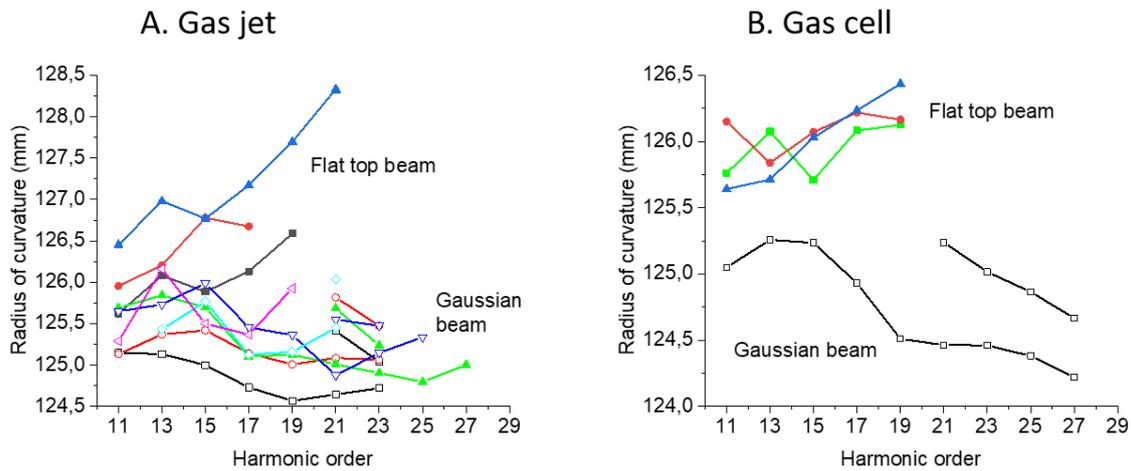

*Fig. S4: Evolution of the wavefront radius of curvature with the harmonic order for harmonics generated with Gaussian beams (open symbols) and with flat-top shaped beams (full symbols) for several experimental runs where the pulse energy and iris diameter are varied. The harmonics are generated in a gas jet (a) or a 1 cm long gas cell (b).*

Analog to Fig. 4 where the radius of curvature and the beam FWHM are shown under the use of a gas jet, figure S5 underlines that the trend is the same for measurements obtained with a 1 cm gas cell.

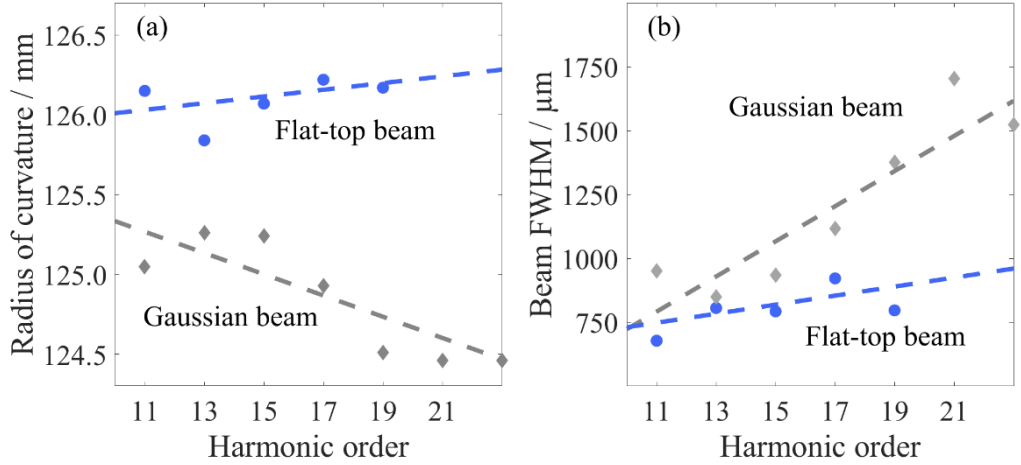

*Fig. S5. (a) Radii of curvature, $R_q$, and (b) beam sizes (FWHM) of the harmonics generated with a Gaussian beam (diamond markers) or with a flat-top shaped beam (circle symbols). Harmonics are generated in a 1cm long gas cell.*

These results provide the waist size and positions when harmonics are generated with a Gaussian driver (Fig. S6 (a) and (c)) or with a flat-top driver (Fig. S6 (b) and (d)). The on-axis harmonic intensities are represented in both cases, and we observe that the foci positions are well separated when harmonics are emitted with a Gaussian beam and less separated when generated with a flat-top beam. Fig. S6 also shows the evolution of the on-axis mean photon energy and bandwidth. As with a gas jet, we observe that the evolution is less pronounced with the flat-top beam than with the Gaussian beam.

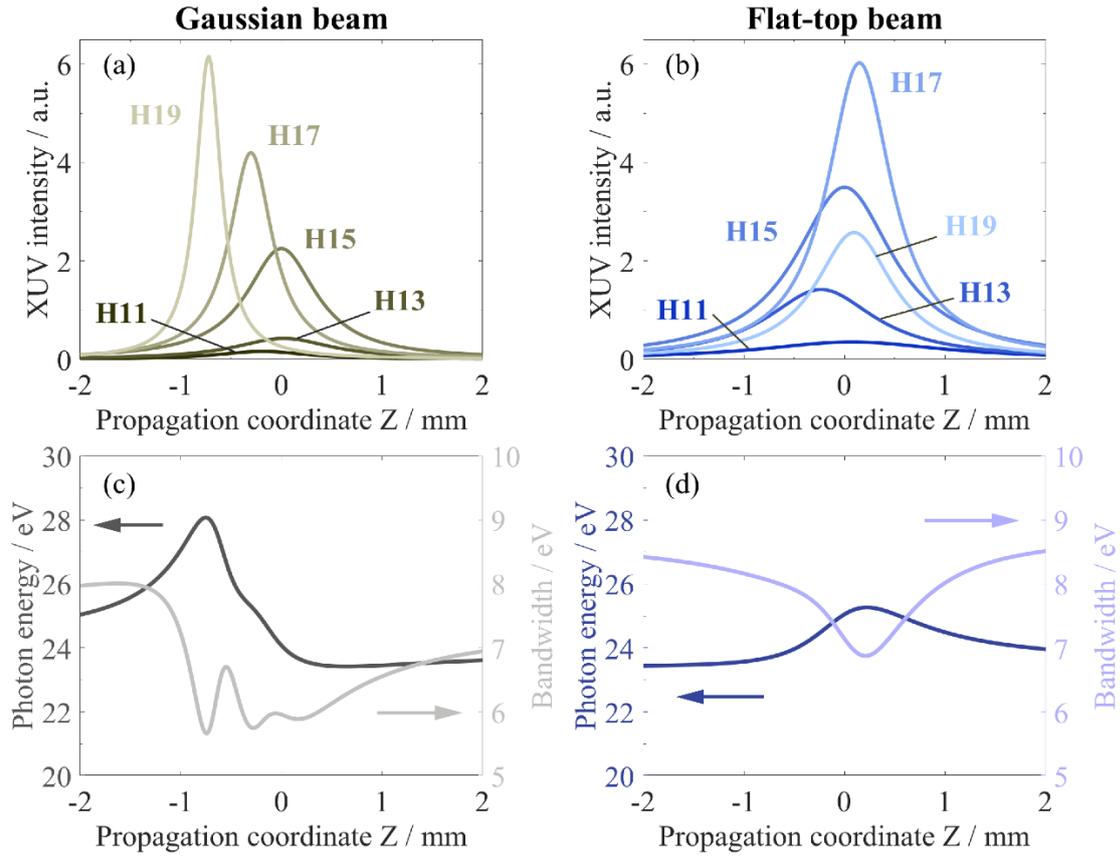

*Fig. S6: Harmonic beam characteristics near XUV focus for harmonics generated with a Gaussian ((a) and (c)), respectively a flat-top shaped ((b) and (d)) fundamental beam in a 1 cm long gas cell (on-axis intensities for harmonics 11 to 19, on-axis mean photon energy and on-axis bandwidth). Z = 0 is arbitrarily set to the position of the focus of harmonic 15.*

Complementary to the results obtained with a gas jet (Fig. 7), Fig. S7 contains the results obtained with a 1 cm gas cell indicating the same difference between a Gaussian fundamental beam and the flat-top shaped driver.

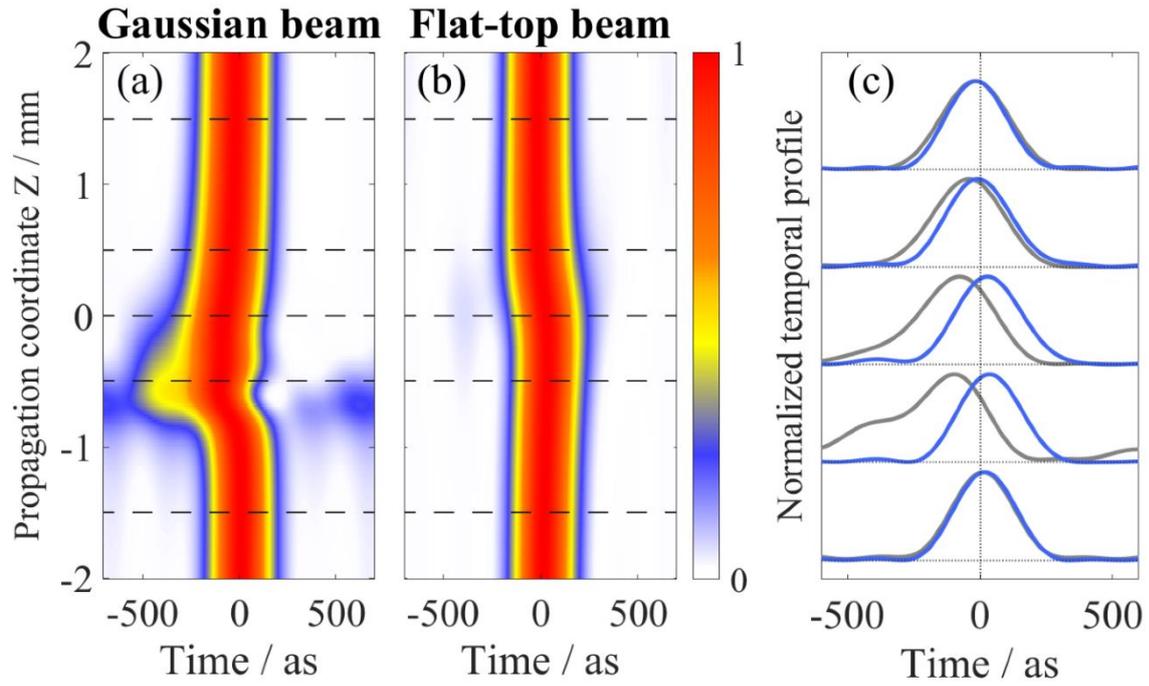

*Fig. S7. Spatial evolution of the attosecond temporal profile for harmonics 11 to 19 emitted in a 1 cm gas cell with a Gaussian fundamental beam (a) or a flat-top shaped fundamental beam (b). The corresponding temporal profiles are shown in subfigure (c) (Gaussian: grey line, Flat-top: blue line) for some longitudinal positions (Z = -1.5 mm, -0.5 mm, 0 mm, 0.5 mm and + 1.5 mm; ordered from bottom to top), as indicated by the dashed line in the temporal profile*

The main paper draws conclusions by considering only the harmonics H11 to H19, generated both with the flat-top and Gaussian beams. Fig. S8 and S9 show the evolution of the attosecond pulse profile with propagation considering all harmonics. We also observe an evolution of the attosecond pulse profile and timing that is more pronounced with the Gaussian beam (S8) than with the flat-top beam (S9).

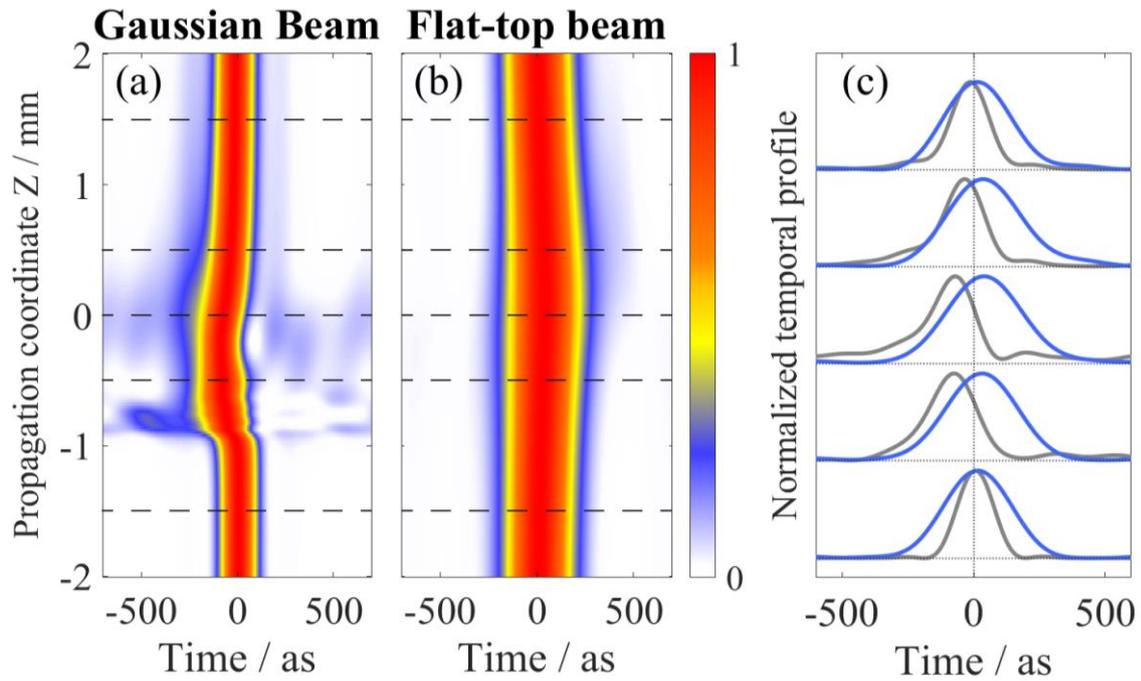

*Fig. S8: Spatial evolution of the temporal profile when all harmonics are considered. Harmonics generated in a gas jet with either a Gaussian (a) or flat-top shaped (b) fundamental beam. (c): Corresponding normalized temporal pulse profile (Gaussian: grey line, Flat-top: blue line) for some longitudinal positions (Z = -1.5 mm, -0.5 mm, 0 mm, 0.5 mm and + 1.5 mm; from bottom to top), indicated as dashed lines in the temporal profile.*

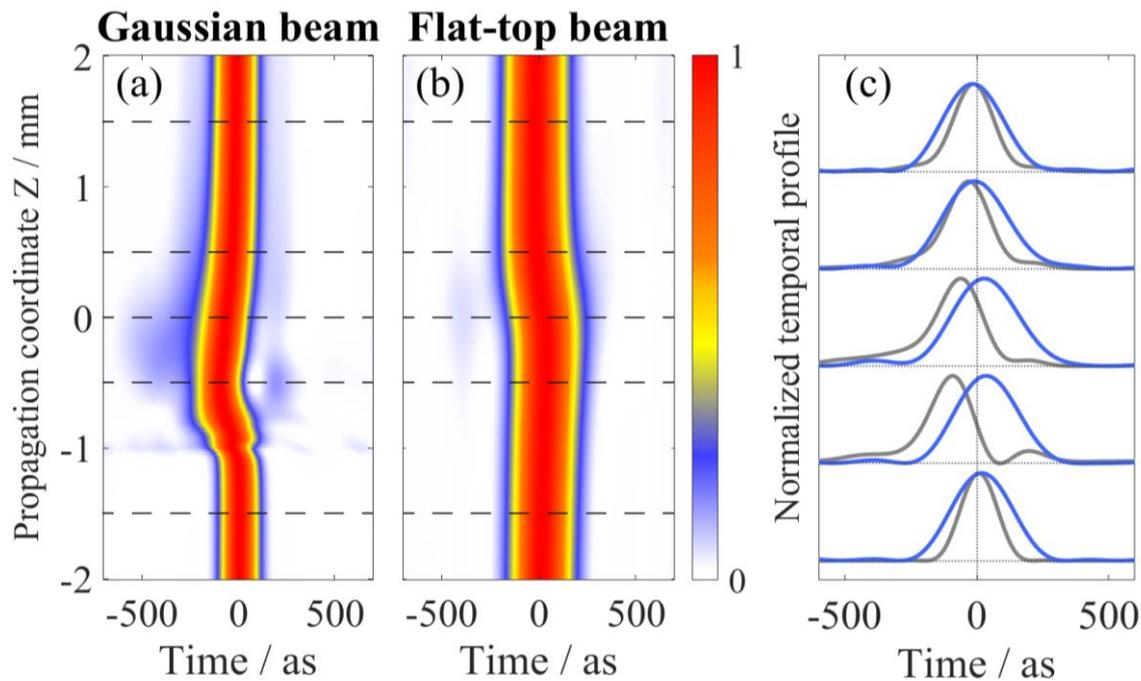

*Fig. S9: Spatial evolution of the temporal profile when all harmonics are considered. Harmonics generated in a 1 cm gas cell with either a Gaussian (a) or flat-top shaped (b) fundamental beam. (c):*

*Normalized temporal pulse profile (Gaussian: grey line, Flat-top: blue line) for some longitudinal positions (Z = -1.5 mm, -0.5 mm, 0 mm, 0.5 mm and + 1.5 mm; from bottom to top), indicated as dashed lines in the temporal profile to the left.*

Fig. S10 (a) shows the harmonic spatial profile measured on the slit plane and the Gaussian fits used in the analysis. The fits are very good for the main part of the peaks. The profile measured on the MCP (Fig. S10 (b)) is also shown with a Gaussian estimate that is not a fit, but an estimate obtained from the sword analysis that considers only measurements on the slit plane. The agreement observed between measured and estimated profiles is good. Harmonics considered in Fig. S10 are generated with a flat-top beam. Similarly, Fig. S11 exhibits harmonics emitted using a Gaussian beam. The agreement between, measurements, fits and estimates are also very good.

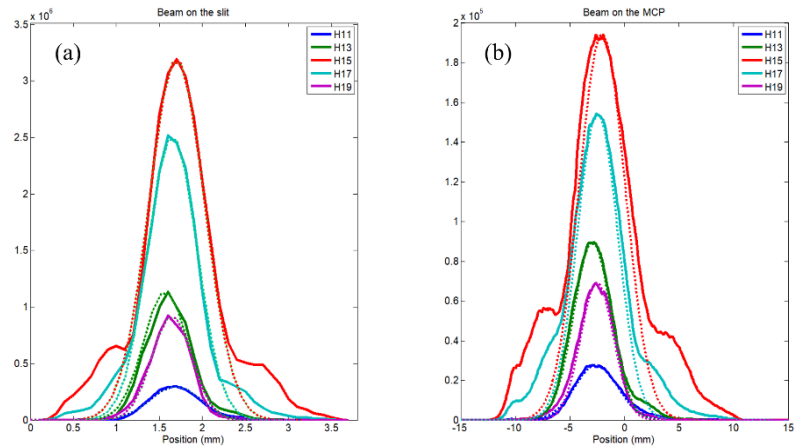

*Fig. S10: XUV profiles obtained with a flat-top IR beam and HHG in a gas jet. (a) Spatial profiles measured in the slit plane and Gaussian fit (dots) used for the Gaussian analysis. (b) Spatial profiles measured on the MCP detector and Gaussian spatial profiles (dots) estimated from the sword analysis relying on measurements in the slit plane.*

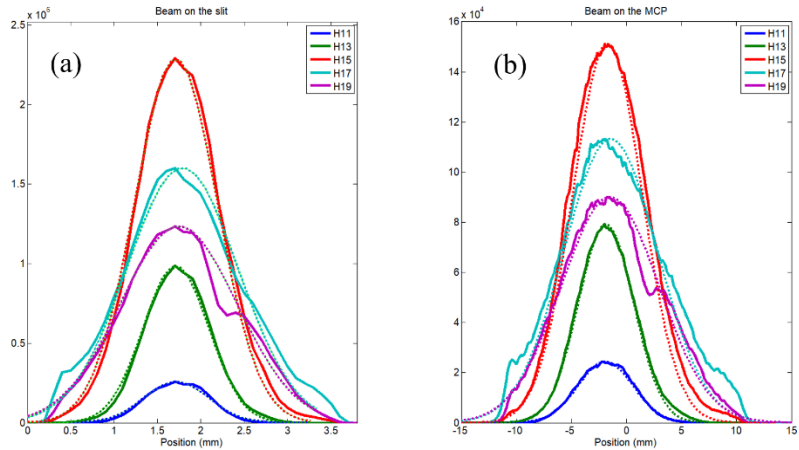

*Fig. S11: XUV Spatial profiles obtained with a Gaussian fundamental beam and HHG in a gas jet. (a) Spatial profiles measured in the slit plane and Gaussian fit (dots) used for the Gaussian analysis. (b)*

*Spatial profiles measured on the MCP detector and Gaussian spatial profiles (dots) estimated from the sword analysis relying on measurements in the slit plane.*

Advanced simulations were performed with TDSE as presented in the text. Fig. S12 shows that the focusing properties are very different for harmonics generated with a Gaussian beam (longitudinally separated foci) compared to a flat-top beam (overlapping foci). Simulations also show that temporal profiles evolve more with propagation in the Gaussian case than in the flat-top case.

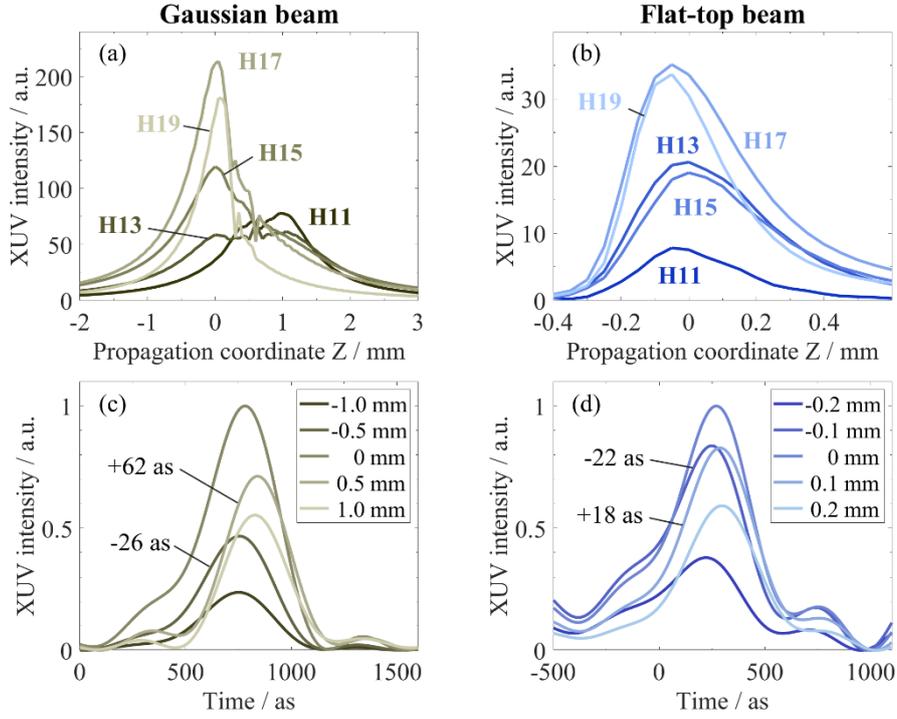

*Fig. S12: XUV-intensity along the propagation axis for harmonic 11 to 19 generated by a IR Gaussian beam (a), and a flat-top respectively (b). Evolution of the attosecond temporal profile for truncated Gaussian (c) and flat-top beams (d) for a peak intensity of $I = 2 \times 10^{14}$ W/cm$^2$. The numbers in (c) and (d) (+62, -26, -22 and +18 as) indicates the shift of the pulse relative to the Z = 0 mm reference.*

For the Gaussian case, the observed increase of the XUV beam size with IR intensity (see Fig. 3. (a)) is not reproduced by the simulations within the approach described above which neglects propagation in the generating medium. To reproduce this evolution, we mimic propagation in a simple way and include an additional dephasing, due to the medium ionization, and the impact of the plasma refractive index. The plasma index evolves radially, affecting the IR beam wavefront. This transverse phase, multiplied by the harmonic order, is transferred to the harmonic beam. This plasma defocusing of the IR leads to corresponding defocusing of the harmonics that is more pronounced with increasing harmonic order. With a gas density that corresponds to an absorption length close to the medium length, we find that plasma defocusing has a strong impact in the Gaussian case (strong intensity gradient and high intensity) but a low impact in the flat-top case (low gradient and low intensity). The simulations consider the plasma defocusing quantitatively reproduce the experimentally observed increase of the XUV beam size with IR intensity and harmonic order. The plasma defocusing also reduces the foci shift by essentially suppressing the XUV light present on axis for Z > 5mm (Fig. S11). The defocusing reduces the evolution of the attosecond pulse duration during propagation but leads to a significant displacement of the center of the atto pulse that can shift by more than 100 as over one mm of longitudinal propagation (Fig. S13).

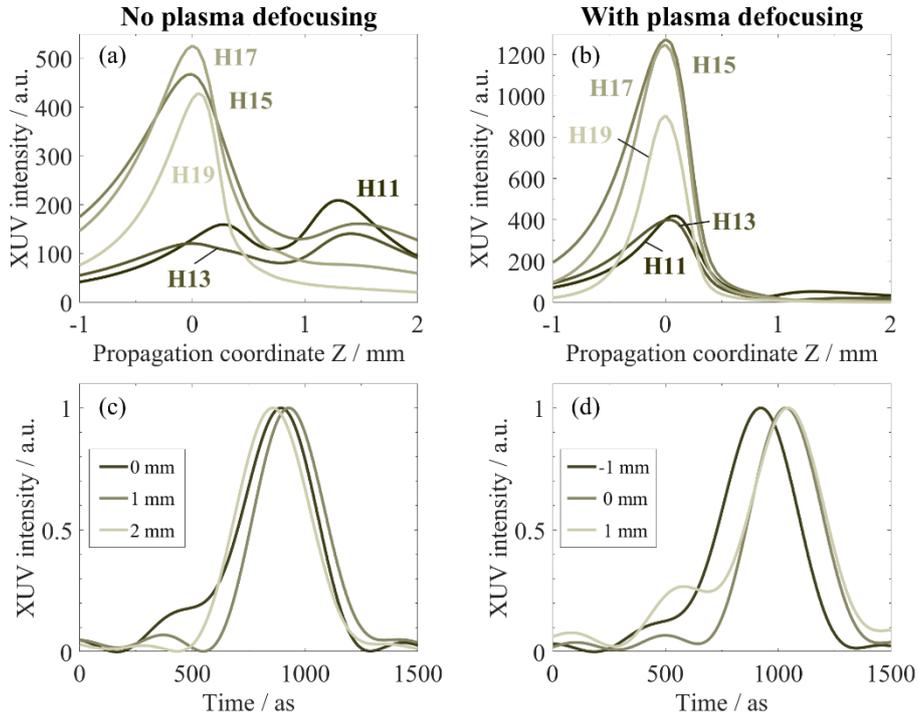

*Fig. S13: On axis harmonics signal for harmonics generated with a truncated Gaussian beam at high intensity (a) without and (b) with the plasma defocusing. Corresponding on axis attosecond profile without (c) and with (d) plasma defocusing at an IR peak intensity of $3.5 \times 10^{14}$ W/cm$^2$.*

For comparison, we simulate the XUV focusing with and without plasma defocusing. The results are shown in Fig. S13. We observe that the plasma defocusing of the IR tends to clean the XUV focusing by essentially suppressing the XUV components that were present at large Z without defocusing. Nevertheless, with the plasma defocusing also, the temporal profile of the attosecond pulses emitted by the Gaussian beam evolves with propagation and the temporal profile near focus differs significantly from the asymptotic pulse profile. There again, the pulse duration changes with propagation and a temporal shift is present as observed experimentally.